\documentstyle[preprint,tighten,aps,epsfig]{revtex}
\begin{document}
\preprint{\vbox{ \null\hfill DFTT 35/98 \\
\null\hfill INFNCA-TH9807 \\
\null\hfill hep-ph/9808426 \\
\vspace{1.0truecm}}}
\draft
\newcommand{\beq}{\begin{equation}}
\newcommand{\eeq}{\end{equation}}
\newcommand{\barr}{\begin{eqnarray}}
\newcommand{\earr}{\end{eqnarray}}
\newcommand{\pup}{p^\uparrow}
\newcommand{\pdown}{p^\downarrow}
\newcommand{\la}{\lambda}
\newcommand{\nd}{\noindent}
\newcommand{\bfk}{\mbox{\boldmath $k$}}
\newcommand{\bfy}{\mbox{\boldmath $y$}}
\newcommand{\sumint}{\mbox{\raisebox{.18ex}{${\scriptstyle\sum}$}
\hspace{-0.47truecm}$\displaystyle\int$}}
\title{Single spin asymmetries in \mbox{\boldmath{$p^{\uparrow}p$}}
and \mbox{\boldmath{$\bar{p}^{\uparrow}p$}} inclusive processes}

\author{ Mauro Anselmino$^{a}$ and Francesco Murgia$^{b}$}
\vskip 12pt
\address{$^{a}$Dipartimento di Fisica Teorica, Universit\`a di Torino\\
and Istituto Nazionale di Fisica Nucleare, Sezione di Torino \\
via P. Giuria 1, I-10125 Torino, Italy \\
$^{b}$Istituto Nazionale di Fisica Nucleare, Sezione di Cagliari\\
and Dipartimento di Fisica, Universit\`a di Cagliari\\
C.P. 170, I-09042 Monserrato (CA), Italy \\}
\maketitle
\begin{abstract}
We consider several single spin asymmetries in inclusive $\pup p$
and $\bar{p}^{\uparrow}p$ processes as higher twist QCD contributions, 
taking into account spin and intrinsic $\bfk_{\perp}$ effects in the 
quark distribution functions. This approach has been previously 
applied to the description of the single spin asymmetries observed in 
$\pup p \to \pi\,X$ reactions and all its parameters fixed: we give 
here predictions for new processes, which agree with experiments for
which data are available, and suggest further possible measurements.  
\end{abstract}

\narrowtext

\section{Introduction}
\label{intro}

In the last years a copious theoretical and experimental activity
has been devoted to the study of single spin asymmetries (SSA) in
inclusive particle production at high energy and moderately large
$p_T$. On theoretical grounds, at leading twist in massless perturbative 
QCD, SSA at high energies and $p_T$ are expected to be negligible.
However, the presently available experimental results seem to show that 
this is not the case: at least in some kinematical regions 
(namely, at large $x_F$ and at moderately large $p_T$) sizeable SSA 
(up to the order of 30\%-40\%) have been measured \cite{exp1,exp2}. These 
experimental results have prompted a renewed theoretical activity, focused 
at the introduction in perturbative QCD schemes of higher twist contributions, 
previously neglected, which could play a crucial role in this context.

Several single spin effects, relevant at different steps of the 
inclusive hadronic production, have been suggested in the literature.
In Ref.s \cite{siv1,siv2,abm}, for example, quark transverse 
momentum effects have been taken into account in the structure of the 
initial, transversely polarized nucleon. A similar mechanism has been 
proposed for the fragmentation process of a polarized quark into the final 
observed particles \cite{col1,artr}. Quark-gluon correlation functions
and gluonic pole contributions have also been considered as possible 
origin of SSA in various processes \cite{ste1,ste2,efre,tery}.
Quark orbital angular momentum and a simple (non QCD) elementary dynamics
is used in Ref. \cite {meng}.

In Ref.~\cite{abm} we start from the QCD formalism for polarized hard 
processes \cite{col2}, with the inclusion of the intrinsic transverse 
momentum of quarks inside the polarized hadrons: this allows to introduce 
a non-diagonal (in the helicity basis) distribution function for these 
quarks, denoted by $\Delta_Nf$. Such distribution 
would be forbidden by time reversal invariance for free quarks \cite{col1}, 
but is allowed by initial state interactions between the incoming hadrons, 
or by quark-gluon correlations \cite{ste1,ste2,efre,tery} or in chiral models 
\cite{drag}. The only assumption is that the QCD factorization theorem 
still holds in such cases. Our formalism, whose results agree with those of 
Sivers \cite{siv1,siv2}, allows then in principle to obtain sizeable values 
for the SSA.

A simple phenomenological parametrization of the new distributions was 
introduced, and all parameters were fixed by fitting the data on SSA in 
inclusive pion production in proton-proton collisions, $\pup p\to\pi\,X$ 
\cite{exp1}. It was shown that a good description of the experimental data 
is indeed possible and the resulting features of the new distribution
turn out to be physically plausible and well justified.

In this paper, equipped with a definite expression for $\Delta_Nf$
obtained by fitting the data on pion production, we further apply our 
formalism to other processes, both obtaining genuine predictions which
can be compared with data already existing and suggesting possible future
measurements and tests. The systematic study of several SSA allows to
evaluate the consistency and relevance of our approach, and to isolate
its contribution to SSA from other possible sources of spin effects.

A preliminary, partial account of this analysis was presented in
\cite{abm2}.

The plan of the paper is the following: in Sec. II we recall and summarize 
our formalism and repeat the fitting procedure which allows to fix all
parameters. Sect. III, which contains the main results of the paper, will 
be devoted to the application of the model to several processes; numerical 
results will be presented and discussed in details. Finally, in Sect. IV we 
give some comments and conclusions.

\section{The model}
\label{model}

In general we shall consider the high energy, high $p_T$, inclusive process 
$h_1^{\uparrow(\downarrow)} h_2\to h_3\,X$, where $h_1$, $h_2$, and
$h_3$ are hadrons; the apex $\uparrow(\downarrow)$ means that
hadron $h_1$ is transversely polarized with respect to the scattering
plane, in the same (opposite) direction as $\mbox{\boldmath $p$}_{h_1}
\times \mbox{\boldmath $p$}_{h_3}$
Some of the hadrons may be substituted by 
other particles, like leptons or photons. In the remaining of this section 
we deal with the process $\pup p\to\pi\,X$; all results can be extended 
to other processes, like those considered in the following section,
by simply adapting the formalism to the case of interest.

The single spin asymmetry $A_N$ for the process under
consideration is defined as follows:

\beq
A_N(x_F,p_T) = \frac{d\sigma^\uparrow-d\sigma^\downarrow}
 {d\sigma^\uparrow+d\sigma^\downarrow}\; ,
\label{an}
\eeq
where $d\sigma^{\uparrow,\downarrow}$ stands for the differential cross section 
$E_\pi \, d^3\sigma^{p^{\uparrow,\downarrow}p \to \pi X} /d^3
\mbox{\boldmath $p$}_{\pi}$; 
$x_F = 2p_L/\sqrt{s}$, where $p_L$ is the pion longitudinal momentum in the 
$p\,p$ c.m. frame, and $\sqrt{s}$ is the total c.m. energy. Of course, other 
sets of kinematical variables could also be used. Single spin asymmetries
for other spin directions are forbidden by parity invariance. 

By allowing for spin effects in the distribution function and still
assuming the QCD factorization theorem to hold, one can write \cite{abm}

\barr
2 \, d\sigma^{unp}A_N(x_F,p_T) &=& 
 \sum_{a,b,c,d}\int d^2\bfk_{\perp_a} \, dx_a \, dx_b \, \frac{1}{\pi z}
 \nonumber \\
 &\times& \Delta^Nf_{a/\pup}(x_a,\bfk_{\perp a}) \> f_{b/p}(x_b) \>
 \frac{d\hat\sigma^{ab\to cd}}{d\hat t} \> D_{\pi/c}(z) \; , 
\label{anqcd}
\earr

\nd where $d\sigma^{unp}$ is the unpolarized differential cross
section for the process under consideration; $f_{b/p}(x_b)$ is
the distribution function, inside the unpolarized proton, of partons $b$  
with a fraction $x_b$ of the proton momentum;
$D_{\pi/c}(z)$ is the unpolarized fragmentation function for parton
$c$ fragmenting into a pion carrying a fraction $z$ of its momentum;
$d\hat\sigma^{ab\to cd}/d\hat t$ is the partonic cross section
for the hard process $ab\to cd$.

The r.h.s. of Eq.~(\ref{anqcd}) differs from the expression of the 
unpolarized cross section $d\sigma^{unp}$ in that the unpolarized distribution 
function of parton $a$ inside the beam proton is replaced by a new, 
nonperturbative function, depending on the intrinsic transverse momentum 
of parton $a$,  

\barr
\Delta^Nf_{a/\pup}(x_a,\bfk_{\perp a}) &=&
 \sum_{\la^{\,}_a}\left[ \hat f_{a, \la^{\,}_a / \pup}
 (x_a,\bfk_{\perp a}) - \hat f_{a, \la^{\,}_a / \pup} (x_a, -\bfk_{\perp a})
 \right] \nonumber \\
 &\equiv& 2\, I^{a/p}_{+-}(x_a,\bfk_{\perp a})\; ,
\label{dn}
\earr
which gives the difference between the total number of partons $a$, with 
momentum fraction $x_a$ and intrinsic transverse momentum $\bfk_{\perp a}$
inside a proton with spin $\uparrow$ and a proton with spin $\downarrow$
[notice that $\hat f_{a, \la^{\,}_a / \pup}(x_a,-\bfk_{\perp a}) = 
\hat f_{a, \la^{\,}_a / \pdown}(x_a,\bfk_{\perp a})$].
This same function is denoted by $f_{1T}^{\perp}$ in Ref. \cite{muld}.
The symbol $I^{a/p}_{+-}(x_a,\bfk_{\perp a})$ has been introduced to show the 
non-diagonal nature, in the helicity basis, of this new quantity \cite{abm}. 

Eq.~(\ref{dn}) explicitely shows that $\Delta^Nf_{a/\pup}(x_a,\bfk_{\perp a})$ 
vanishes when $\bfk_{\perp a} \to 0$. Parity invariance requires $\Delta^Nf$
to vanish when $\bfk_{\perp a}$ is perpendicular to the scattering plane
({\it i.e.} parallel to the proton spin), so that
\beq
\Delta^Nf_{a/\pup}(x_a,\bfk_{\perp a}) =
\hat f_{a/\pup} (x_a,\bfk_{\perp a}) - \hat f_{a/\pdown} (x_a, \bfk_{\perp a})
\sim k_{\perp a} \, \sin\varphi
\label{sinp}
\eeq
where $\varphi$ is the angle between $\bfk_{\perp a}$ and the $\uparrow$ 
direction (normal to the scattering plane).

We notice that $\Delta^Nf_{a/\pup}(x_a,\bfk_{\perp a})$
is an odd function of $\bfk_{\perp a}$. This means that we cannot
neglect the $\bfk_{\perp a}$ dependence of all the other terms in
the convolution integral, Eq.~(\ref{anqcd}). We have to keep into 
account $\bfk_{\perp a}$ values in the partonic cross sections
and the differences of dynamical contributions from partons with 
an intrinsic $+\bfk_{\perp a}$ and those with an intrinsic $-\bfk_{\perp a}$,
which leads to an overall value of $A_N$ due to higher twist effects. 

In fact Eq.~(\ref{anqcd}) for $A_N$ can be rewritten as:

\barr
2\,d\sigma^{unp}A_N(x_F,p_T) &=& 
 \sum_{a,b,c,d}\int d^2\bfk_{\perp_a} \, dx_a \, dx_b \, \frac{1}{\pi z}
 \Delta^Nf_{a/\pup}(x_a,\bfk_{\perp a}) \> f_{b/p}(x_b)
 \nonumber \\
 &\times&  \left[ \frac{d\hat\sigma^{ab\to cd}}{d\hat t}(\bfk_{\perp_a}) -
 \frac{d\hat\sigma^{ab\to cd}}{d\hat t}(-\bfk_{\perp_a}) \right] \>
 D_{\pi/c}(z)\; , 
\label{ank}
\earr

\nd where now the integration on $\bfk_{\perp_a}$ runs only over
the positive half-plane of its components. 
The momentum fraction $z$ is fixed in terms of $x_a$, $x_b$ and
$\bfk_{\perp a}$ by momentum conservation in the partonic process.

The only unknown function in Eq.~(\ref{ank}) is the distribution 
$\Delta^Nf_{a/\pup}(x_a,\bfk_{\perp a})$, which, at least in principle, 
is then measurable via the single spin asymmetry $A_N$.

In order to perform numerical calculations, we make some
further assumptions which, while preserving the basic physical ideas
of our approach, make computations easier to handle.
Our first assumption is that the dominant effect is given by the valence 
quarks in the polarized protons. That is, we assume  
$\Delta^Nf_{a/\pup}(x_a,\bfk_{\perp a})$ to be non-zero 
only for valence $u$, $d$ quarks: while it is natural to assume a correlation
between the proton polarization and the $\bfk_{\perp}$ of the polarized 
valence quarks, one does not expect so for the sea quarks. Moreover, 
sea quarks do not contribute much to the production of large $x_F$ pions. 
Secondly, we evaluate Eq. (\ref{ank}) by assuming that the main 
contribution comes from $\bfk_{\perp a} = \bfk_{\perp a}^0$ where,
as suggested by Eq. (\ref{sinp}), $\bfk_{\perp a}^0$ lies in the overall 
scattering plane and its magnitude equals the average value of 
$\langle \bfk^2_{\perp a}\rangle^{1/2}= k_{\perp_a}^0(x_a)$. This average 
value -- which sets an overall physical scale for the transverse momentum 
effects -- will in general depend on $x_a$. Estimates of this dependence 
have been given in the literature \cite{gros}; it can be well represented 
by the following expression ($M$ being the proton mass):
\beq
\frac{1}{M}k_{\perp a}^0(x_a) =
 0.47 \> x_a^{0.68}(1-x_a)^{0.48} \,.
\label{kx}
\eeq

The residual $x_a$ dependence in $\Delta^Nf_{a/\pup}$ not coming from
$k_{\perp a}^0$ is taken to be of the simple form 

\beq
N_a x_a^{\alpha_a}(1-x_a)^{\beta_a} \,,
\label{ix}
\eeq
where $N_a$, $\alpha_a$ and $\beta_a$ are free parameters.

We then end up with the simple expression:
\barr
&&\int d^2\bfk_{\perp_a} \Delta^Nf_{a/\pup}(x_a,\bfk_{\perp a}) 
 \left[ \frac{d\hat\sigma^{ab\to cd}}{d\hat t}(\bfk_{\perp_a}) -
 \frac{d\hat\sigma^{ab\to cd}}{d\hat t}(-\bfk_{\perp_a}) \right] \nonumber \\
&\simeq& \frac{k_{\perp_a}^0(x_a)}{M} \> N_a x_a^{\alpha_a}(1-x_a)^{\beta_a}
 \left[ \frac{d\hat\sigma^{ab\to cd}}{d\hat t}(\bfk_{\perp_a}^0) -
 \frac{d\hat\sigma^{ab\to cd}}{d\hat t}(-\bfk_{\perp_a}^0) \right] \,.
\label{app}
\earr
where $k_{\perp_a}^0(x_a)$ is given by Eq. (\ref{kx}) and, choosing 
$xz$ as the scattering plane and $z$ as the direction of the incoming
polarized proton, $\bfk_{\perp_a}^0 = (k_{\perp_a}^0,0,0)$.

Eqs. (\ref{ank}), (\ref{kx}) and (\ref{app}) can be used
to explain measured single spin asymmetries and to give numerical 
predictions for other single spin asymmetries of interest,
as we shall do in the sequel. Parametrizations for the unpolarized
partonic distributions are available in the literature; similarly for 
the fragmentation functions, for which, due to the increasing experimental 
information available, a lot of progress has been recently made.
Finally, the analytical expressions of the elementary cross sections 
for all possible $ab\to cd$ partonic processes are well known; we only need 
to take into account the corresponding kinematical modifications due to the 
inclusion of the transverse momentum for parton $a$.

In Ref. \cite{abm} the parameters appearing in the expression of 
$\Delta^Nf_{a/\pup}(x_a,\bfk_{\perp a})$, Eqs. (\ref{ix})
and (\ref{app}), were fixed by fitting the available data on single spin 
asymmetries for the $\pup p\to \pi\,X$ process \cite{exp1}.
The main result of Ref. \cite{abm} was indeed to show that it is possible 
within such an approach to reproduce the experimental data with physically 
reasonable values of the parameters.

\begin{figure}
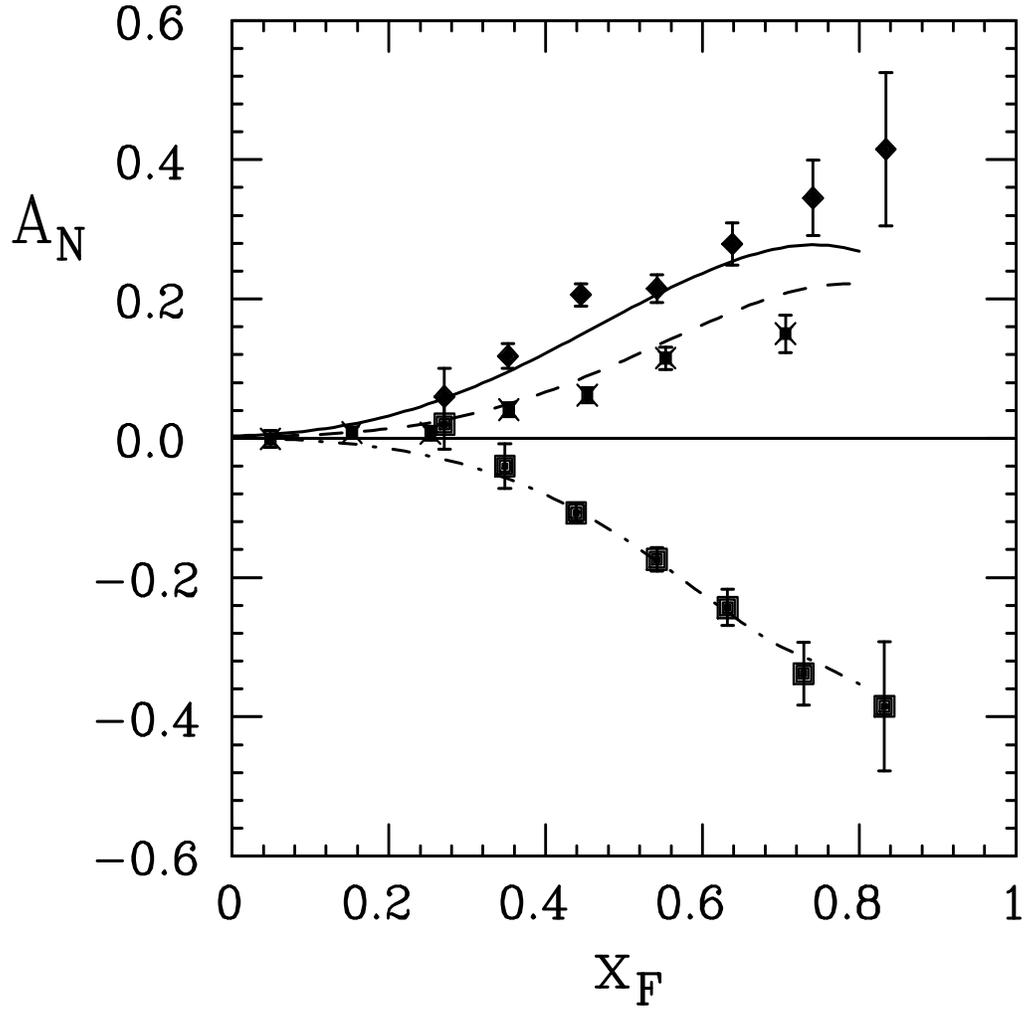

\caption{Fit of the data on $A_N$ for the process $p^\uparrow p\to\pi\,X$
\protect\cite{exp1}, with the parameters given in Eq.~(\protect\ref{par});
the upper, middle, and lower sets of data and curves refer respectively
to $\pi^+$, $\pi^0$, and $\pi^-$. }
\label{pp}
\end{figure}

The precise values of the parameters depend on the specific choice adopted 
for the distribution and fragmentation functions. We have repeated here 
the fitting procedure of Ref. \cite{abm} choosing the MRSG parametrization  
\cite{mrsg} for the partonic distribution functions, and the parametrization 
(BKK1) of Ref. \cite{bkk1} for pion and kaon 
fragmentation functions. The quality of the fit -- see Fig. \ref{pp} -- 
is quite similar to the original one, while the values of the parameters 
do not change significantly:
\begin{equation}
\begin{array}{lccc}
\rule[-0.6cm]{0cm}{1.3cm}
 \; & \;\; N_a \;\; & \;\; \alpha_a \;\; & \;\; \beta_a \\
\rule[-0.6cm]{0cm}{0.5cm}
 u \;\;\;\;\;   & \;\; 3.68 \;\; & \;\; 1.34 \;\;  &  \;\; 3.58 \\
\rule[-0.6cm]{0cm}{0.5cm}
 d \;\;\;\;\;   & \!\!\!\!\!\;\; -1.24 \;\; & \;\; 0.76 \;\; & \;\; 4.14 \\
\end{array}
\label{par}
\end{equation}



Having fixed the values of the parameters we can now give predictions
for several other single spin asymmetries, some of which have been
measured; this is the main purpose of the paper. 

\section{New results}
\label{res}

In this section we consider a number of interesting physical processes.
The full set of such processes and the corresponding experimental data,
either available now or, hopefully, in the near future, should allow to 
assess the validity of our model. Unless explicitely stated, we use the 
formalism, the sets of distribution and fragmentation functions discussed
and the parameters derived in the previous section.

\subsection{$\protect\bbox{\bar{p}^\uparrow\!\! p \to \pi\,X}$}
\label{pb}

Recently the E704 Collaboration at Fermilab has presented results on SSA 
for inclusive production of pions in the collision of transversely polarized 
antiprotons off a proton target \cite{exp2}. The kinematical conditions are 
the same as in the case of polarized protons \cite{exp1}. Within our model 
the connection between SSA with polarized protons or antiprotons is very 
simple: since only valence quark contributions are taken into account, we
just have to exploit charge conjugation relations for the $I_{+-}$ 
distributions, {\it i.e.} $I^{\bar u/ \bar p}_{+-} = I^{u/p}_{+-}$
and $I^{\bar d/\bar p}_{+-} = I^{d/p}_{+-}$. In particular, this means that 
\beq
A_N(\bar{p}^\uparrow p\to\pi^\pm) \simeq A_N(\pup p\to\pi^\mp)
\quad\quad\quad A_N(\bar{p}^\uparrow p\to\pi^0)
\simeq A_N(\pup p\to\pi^0) \,.
\label{resppbar}
\eeq

\begin{figure}
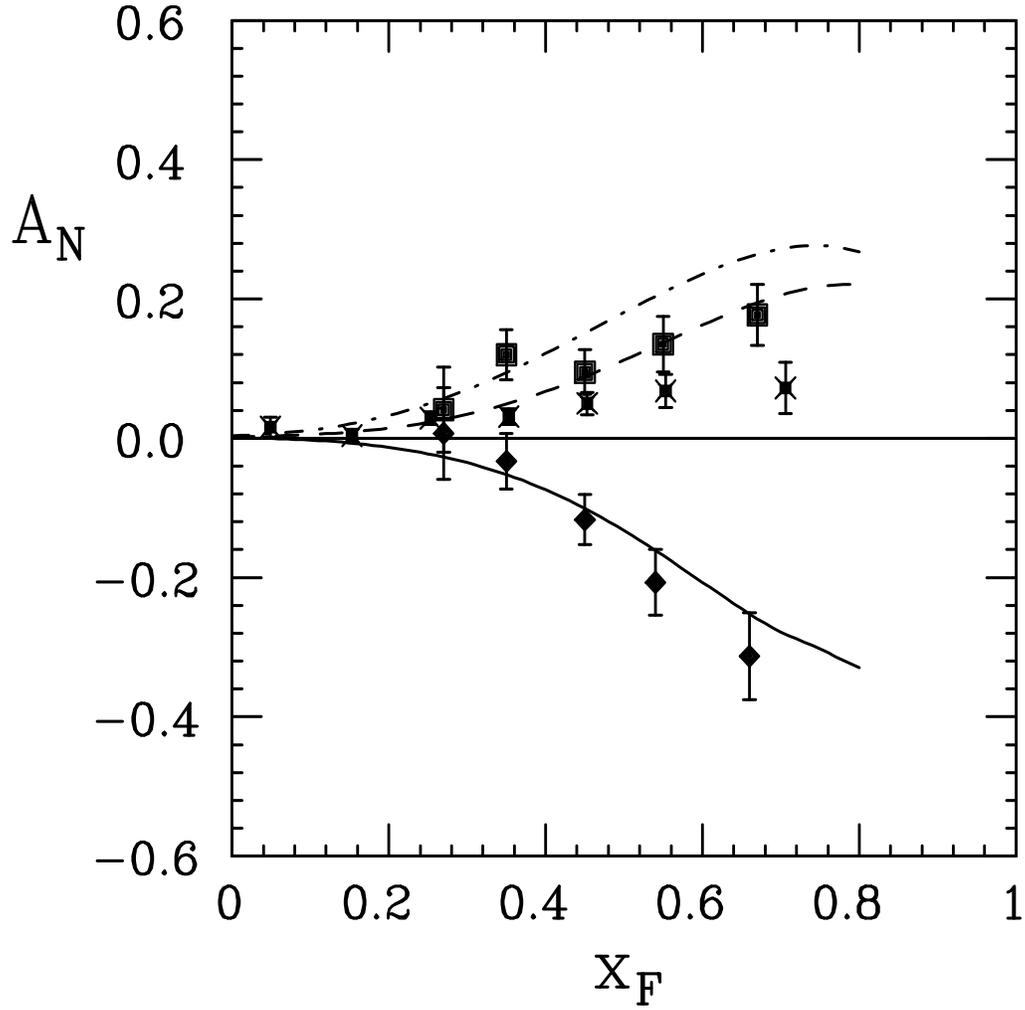

\caption{ Single spin asymmetries $A_N$ for the process
$\bar{p}^\uparrow p\to\pi\,X$ \protect\cite{exp2}; the lower, middle,
and upper sets of data and curves refer respectively
to $\pi^+$, $\pi^0$, and $\pi^-$. }
\label{pbar}
\end{figure}

As it is shown in Fig. \ref{pbar}, this compares rather well with experimental
results, although our results for $\pi^0$ and $\pi^-$ are somewhat too large.

\subsection{$\protect\bbox{\pup\!\!(\bar p^\uparrow\!\!)\,p\to \pi\,X}$
at fixed $\protect\bbox{x_F}$, as a function of $\protect\bbox{p_T}$}
\label{pt}

\begin{figure}
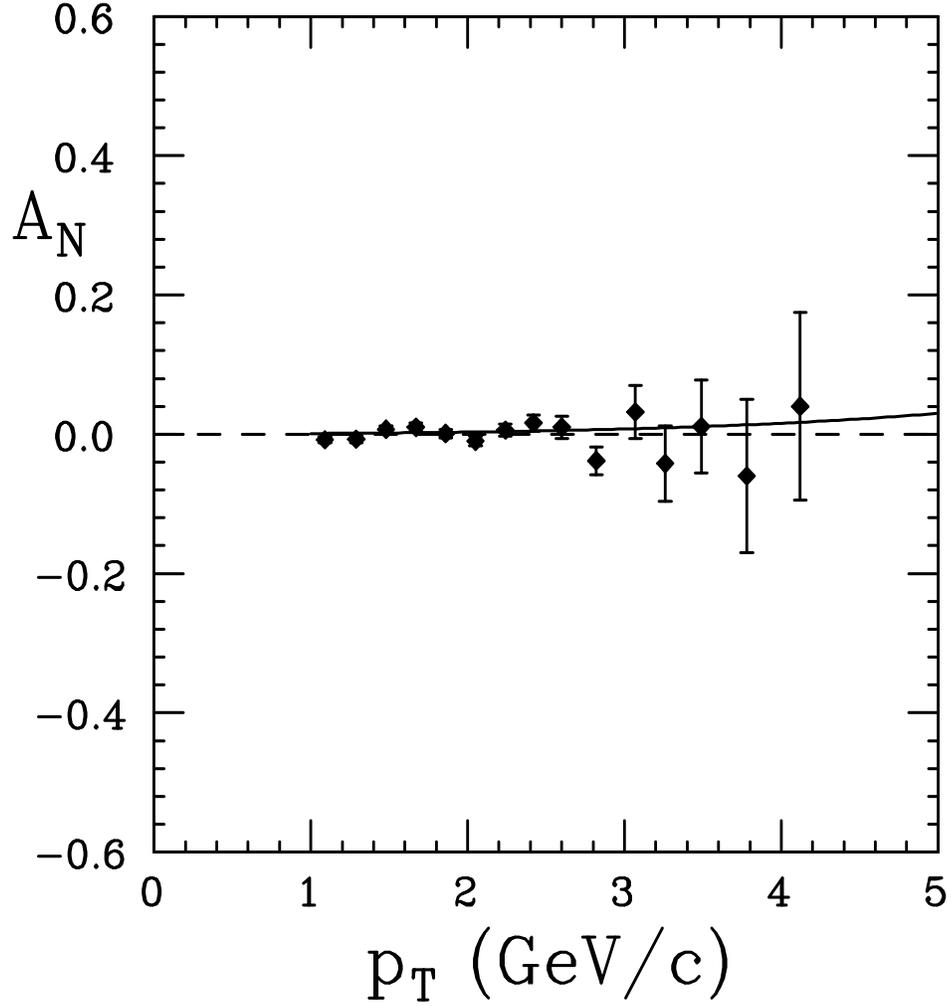

\caption{ Single spin asymmetry for the process
$p^\uparrow p\to\pi^0\,X$ at fixed $x_F$, as a function of $p_T$;
experimental data, at $|x_F| \leq 0.15$, are from Ref.~\protect\cite{exp3};
the solid curve shows our corresponding theoretical prediction at $x_F=0$.}
\label{xf0}
\end{figure}

From presently available experimental data it is not possible to disentangle 
the behaviour of $A_N(x_F,p_T)$ as a function of $x_F$ and $p_T$ separately. 
This might be crucial for the refinement of theoretical models.
The E704 Collaboration has published results on SSA for the processes 
$\pup(\bar{p}^\uparrow) p \to \pi^0\,X$ for $|x_F|\le 0.15$, in the $p_T$ range
$1.0-4.5$ GeV/$c$ \cite{exp3}. No sizeable value of $A_N$ was measured. 
The results of our model are in good agreement with these data: we show them 
in Fig. \ref{xf0} for $\pup p \to \pi^0\,X$ (a similar agreement holds
for antiprotons). Only at values of $x_F$ greater than $\sim 0.3$ 
sizeable effects for the SSA can be obtained.

\subsection{$\protect\bbox{\pup\!\!(\bar p^\uparrow\!\!)\,p\to \gamma\,X}$}
\label{gamma}

Unfortunately, only few experimental data \cite{exp4}, at $x_F\sim 0$
and $p_T$ in the range $2.5-3.1$ GeV/$c$
are available for this process at present.
We stress its importance: possible origins of SSA arising  
in final quark fragmentation \cite{col1,artr} are obviously excluded here; 
in principle we could then directly test those effects originating in  
initial state interactions and/or from quark-gluon correlations. 

\begin{figure}
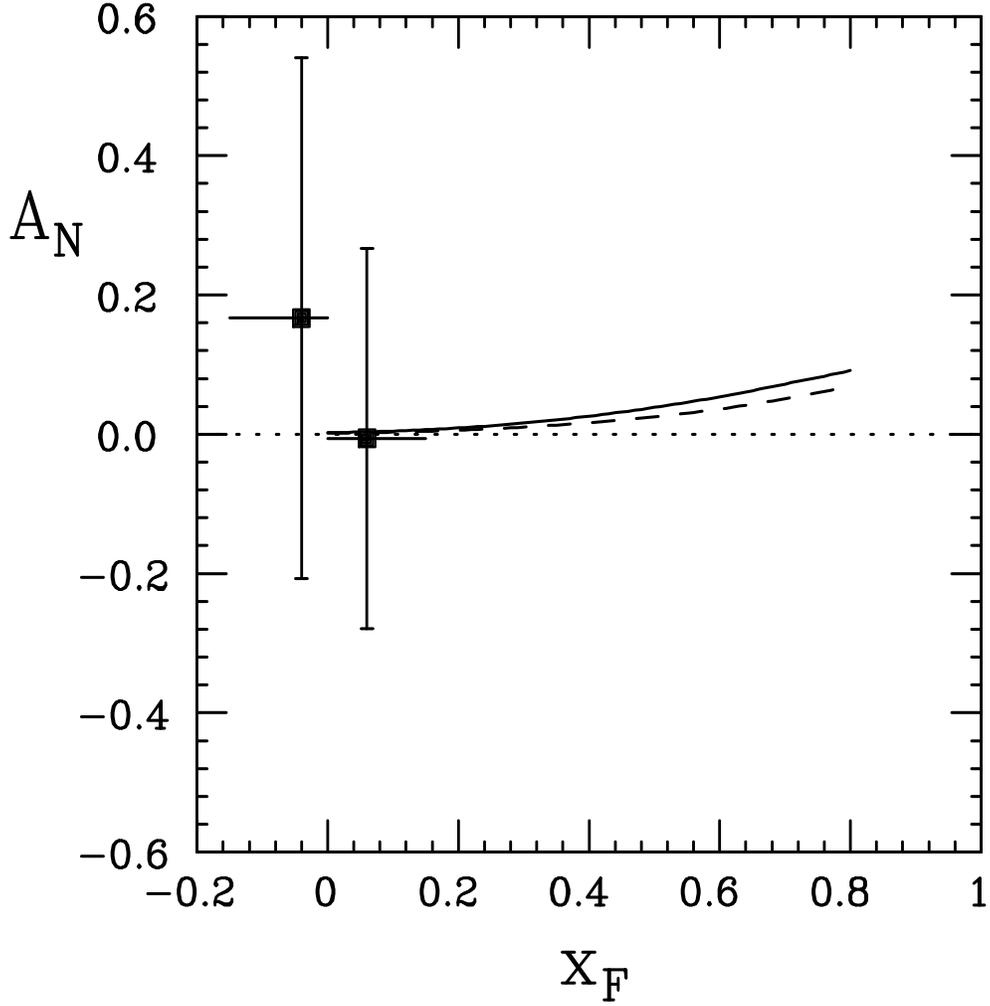

\caption{ Single spin asymmetry for the process 
$p^\uparrow (\bar{p}^\uparrow) \, p\to\gamma\,X$; experimental data,
at $|x_F| \leq 0.15$ and $2.5 < p_T < 3.1$ GeV/$c$,
are from Ref.~\protect\cite{exp4}. The curves show our corresponding
theoretical predictions at $p_T=2.5$ GeV/$c$; the solid curve
refers to the $p^\uparrow p\to\gamma\,X$ process, the dashed curve to
the $\bar{p}^\uparrow p\to\gamma\,X$ case. }
\label{gm}
\end{figure}

The predictions of our model are presented in Fig. \ref{gm}, as a function
of $x_F$. We choose the same kinematical conditions as in the available 
experimental data of E704 Collaboration \cite{exp4}. Only two experimental 
points at $x_F \sim 0$, with huge errors, are available; they are in agreement
with our results, but, given the large errors bars, we do not consider such 
an agreement as significant; further experimental data would be most helpful.

Theoretical estimates for SSA in $\pup p\to \gamma\,X$ were given also by 
Qiu and Sterman \cite{ste1}, who presented predictions for kinematical 
conditions similar to those of E704 collaboration \cite{exp4}.
Their theoretical model introduces higher twist distribution functions 
arising from soft quark-gluon correlations: results are given for two possible 
choices of these unknown higher twist distributions. 
These results have a very similar behaviour, as a function of $x_F$, when 
compared with ours, Fig. \ref{gm}. However, they are greater by approximately 
a factor 2.

In Ref. \cite{shaf} it has been argued that out of the two parametrizations 
for the higher twist distributions proposed in Ref. \cite{ste1} the first one 
(which produces the larger result for the asymmetry) should be disregarded, 
since it violates some required physical constraints. The second one should 
be multiplied by an overall factor smaller that unity, leading to results on 
SSA reduced by at least a factor two.
These considerations, if correct, reduce considerably the difference
between the results of Qiu and Sterman and our present results.

\subsection{$\protect\bbox{\pup\!\!(\bar p^\uparrow\!\!)\,p\to K\,X}$}
\label{kappa}

It is interesting to investigate the inclusive production of hadrons 
different from the pion. On one hand, in fact, such new data would 
allow further tests of our model and of the parametrization of the
$I_{+-}$ distributions; on the other hand, the study of kaon production, 
for example, is interesting by itself, due to the presence of valence $s$ 
quarks and the possibility of learning about the role of strange quarks in
fragmentation processes.

To study kaon production, we need information on the corresponding 
fragmentation functions $D_{K/c}$. The knowledge of kaon FF is more 
limited than in the pion case and, to the best of our knowledge, only a few 
parametrizations are available in the literature. Let us briefly recall the 
main features of these parametrizations, which will be used in the sequel:

\noindent i)
Ref. \cite{bkk1} not only gives the set of fragmentation functions 
(BKK1) used to fit the pion data, but also gives parton fragmentation 
functions into kaons. Although not the most recent one, it has the advantage 
to give separate contributions from leading (valence) and non-leading (sea) 
parent quarks of the produced meson, which allow to derive separate FF's 
for positively and negatively charged mesons;

\noindent ii)
The same authors have recently published a more accurate 
parametrization (BKK2) \cite{bkk2}, based on a much richer set of experimental 
data. However, this second set only gives FF either for the sum of charged 
mesons ($\pi^+ + \pi^-$, $K^+ + K^-$) or for neutral 
($\pi^0$, $K^0 + \bar{K}^0$) mesons. 

\noindent iii)
Other sets of parametrizations for the sum of charged pion and kaon FF have
been given by Greco and Rolli (GR) \cite{gr1,gr2}. 

\noindent iv)
Finally, some most recent fragmentation functions (IMR) can be found in 
Ref. \cite{imr}. 

Since the main contribution to the large $x_F$ production of mesons 
comes from partonic processes involving valence quarks from the initial 
nucleon and in the final mesons, it is plausible to expect, for kaons
produced from polarized protons, the following behaviour, in analogy to 
the pion case:

\beq
  A_N(K^+)\sim A_N(\pi^+)\,;\qquad A_N(K^0)\sim A_N(\pi^-)\,;\qquad
  A_N(K^-)\sim A_N(\bar{K}^0) \sim 0 \;.
  \label{ak}
\eeq

This is clearly understood if we keep in mind the flavour content
of pions and kaons ($\pi^+=u\bar{d}$, $\pi^-=d\bar{u}$,
$\pi^0=(u\bar{u}-d\bar{d})/\sqrt{2}$; $K^+=u\bar{s}$, $K^-=s\bar{u}$,
$K^0=d\bar{s}$, $\bar{K}^0=s\bar{d}$) and remember that we are
assuming that only $I_{+-}^{u,d}$ are different from zero for the
incident polarized proton.

Regarding $K^0_S$ mesons, which are actually observed in experiments,
since $\sigma(K^0_S) = \bigl[\sigma(K^0)+\sigma(\bar{K}^0)\bigr]/2$,
using Eq.~(\ref{an}) we see that the asymmetry for $K^0_S$ is related
to the asymmetries for $K^0$ and $\bar{K}^0$ by: 

\beq
  A_N(K^0_S) = \frac{A_N(K^0) + \frac{\displaystyle \sigma^{unp}(\bar{K}^0)}
  {\displaystyle \sigma^{unp}(K^0)}A_N(\bar{K}^0)}
  {1+\frac{\displaystyle \sigma^{unp}(\bar{K}^0)}
  {\displaystyle \sigma^{unp}(K^0)}} \> \cdot
  \label{aks}
\eeq

\begin{figure}
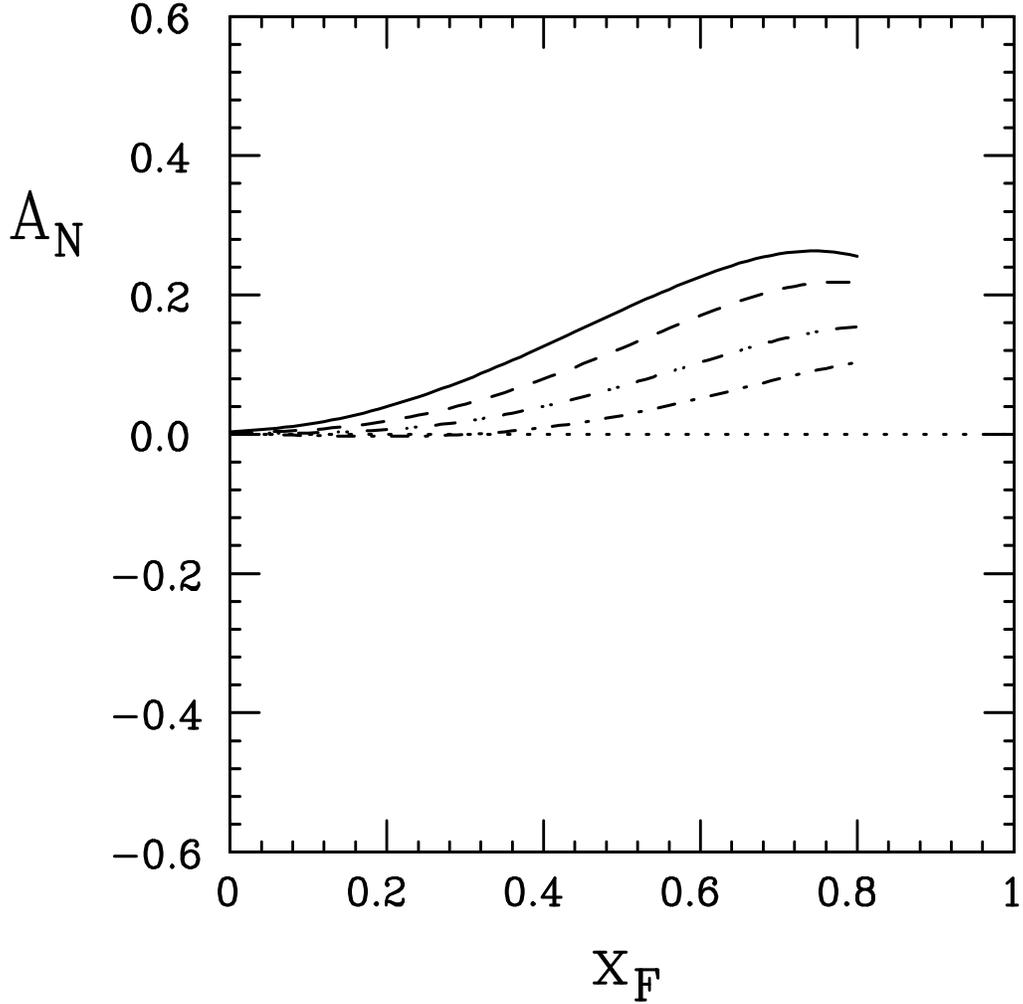

\caption{
Predicted single spin asymmetries for the process $p^\uparrow p\to K\,X$,
with the set of kaon FF's BKK1 \protect\cite{bkk1}; kinematical conditions
are the same as for the pion case, at $p_T=1.5$ GeV/$c$. The solid,
dashed, dot-dashed, double dot-dashed curves refer respectively to
the $K^+$, $K^-$, $K^0$, $K^0_S$ cases. Results for $\bar{K}^0$ meson
are very similar to those for $K^-$ case. }
\label{k1}
\end{figure}

Since we have used the BKK1 set of FF's for fitting the pion data,
the first thing we can do is to use the corresponding BKK1 FF's for
the kaons, keeping the same $I^{u,d}_{+-}$ functions as obtained from 
fitting the pion asymmetry data. Then, by inserting the $D_{K/c}$ given in
Ref. \cite{bkk1} into Eq. (\ref{ank}) and using Eqs. (\ref{kx}), (\ref{app})
and (\ref{par}), we obtain predictions for kaon SSA, in the same kinematical
region as those observed for pions \cite{exp1,exp2}: they are shown in
Fig.~\ref{k1}.

\begin{figure}
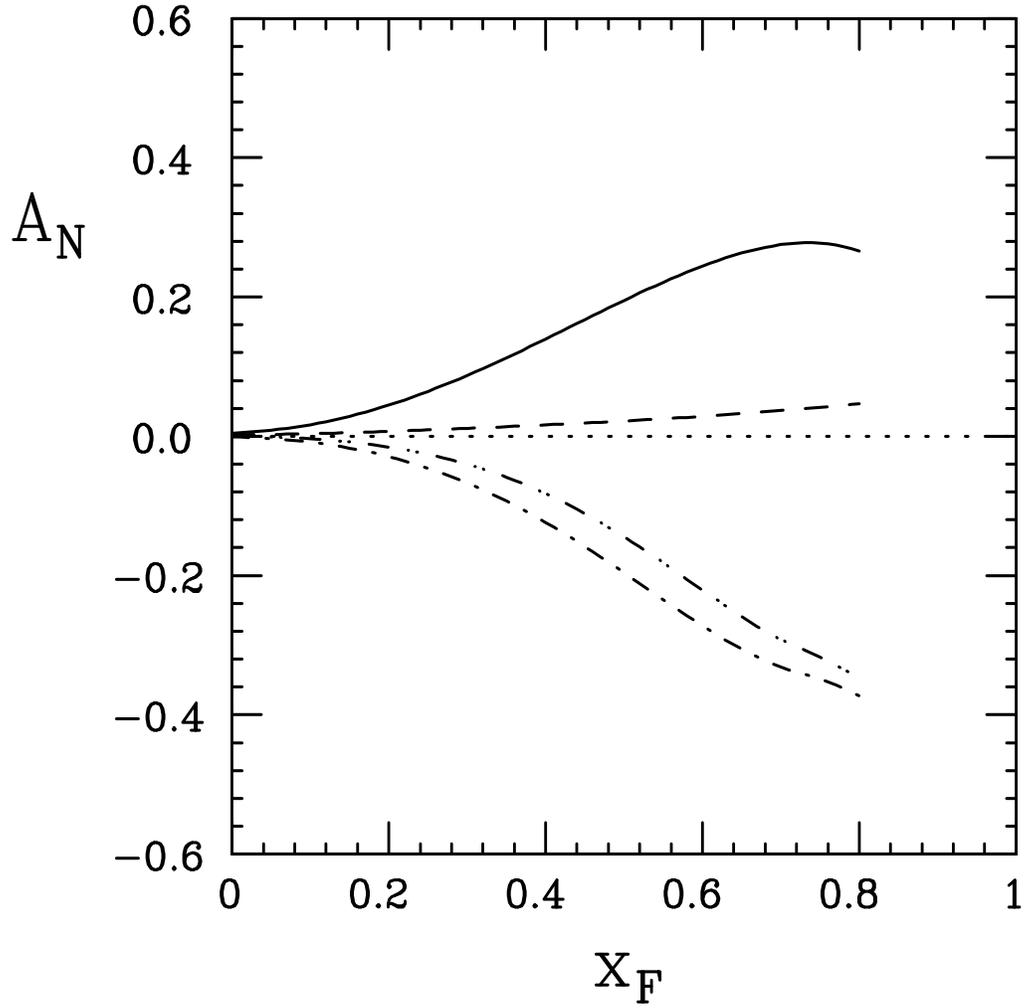

\caption{
The same as for Fig.~\protect\ref{k1}, but using the set of kaon FF's
BKK1 modified so that $D_{K/sea}\simeq D_{K/val}[D_{\pi/sea}/D_{\pi/val}]$
(see text for more details). }
\label{kscaled}
\end{figure}

Surprisingly, the results for $K^-$ and $K^0$ 
are very different from what expected qualitatively, 
see Eq.~(\ref{ak}) and compare Figs.~\ref{pp} and \ref{k1}.
A look at the parametrization of the BKK1 FF's, $D_{K/q}(z)$, for the leading
(valence) parent quarks ({\it e.g.} $u$ quark inside $K^+$) and the 
non-leading (sea) ones ({\it e.g.} $s$ quark inside $K^+$) helps to understand 
the origin of such a discrepancy: in BKK1 FF the non-leading contributions are 
not suppressed, for high values of $z$, compared to the leading ones, which is 
what one would have expected on general grounds, leading to Eq.~(\ref{ak}).
That the unexpected behaviour of Fig.~\ref{k1} originates from this property 
of the BKK1 FF can be confirmed in a simple way: if we take the valence 
contributions of BKK1 kaon FF's, but {\it rescale} the sea ones by assuming 
the same relative behaviour as in the BKK1 pion case
(that is, we redefine $D_{K/sea} \simeq D_{K/val} [D_{\pi/sea}/D_{\pi/val}]$), 
then the results for kaon asymmetries agree very well
(see Fig.~\ref{kscaled}) with the expected qualitative behaviour, 
Eq.~(\ref{ak}). Of course, this simple example has no 
physical motivation, but only clarifies the reason why kaon asymmetries
in Fig.~\ref{k1} are so different from what expected. The real, physical
origin of this discrepancy is in the behaviour at high $z$
of non-leading {\it vs.} leading contributions in the BKK1 FF's.
Indeed, results consistent with Eq. (\ref{ak}) are found -- as we 
shall show -- when using fragmentation functions for which the sea quark 
contribution is suppressed with respect to the valence one. 

\begin{figure}
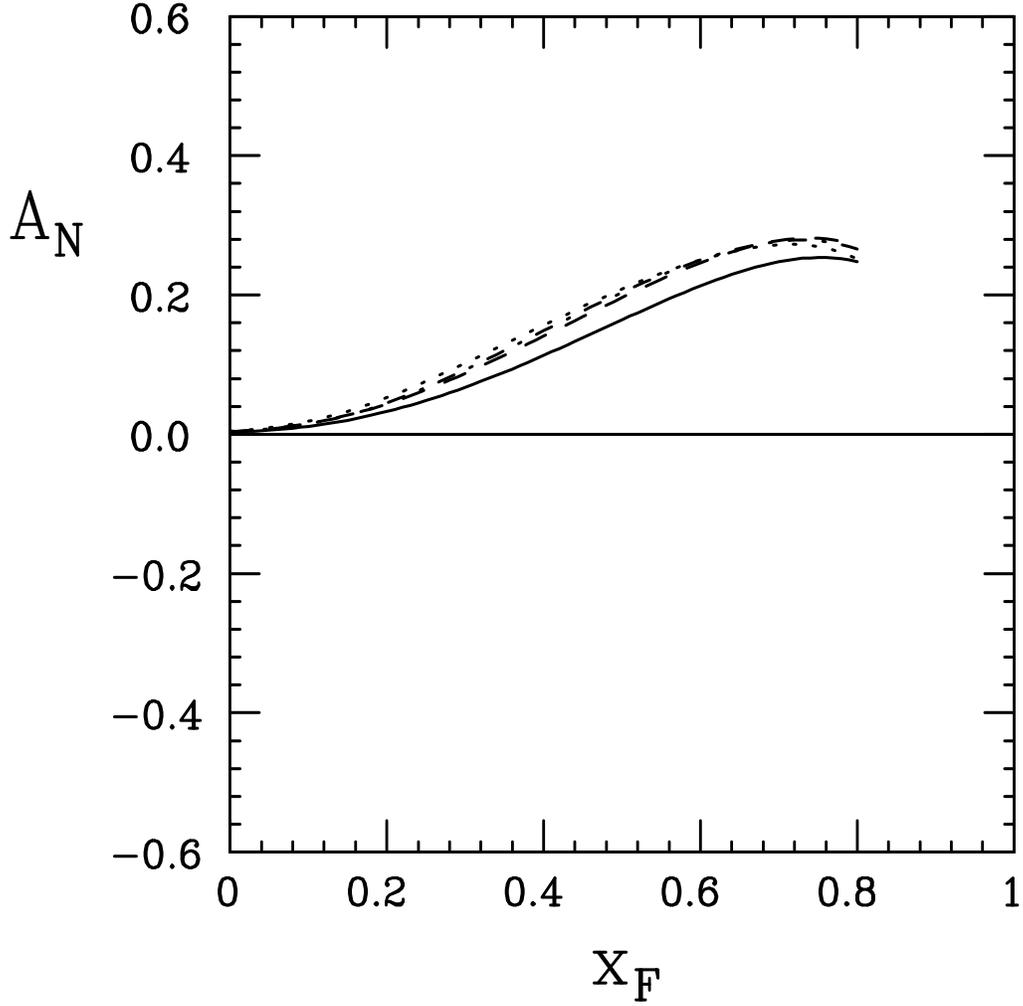

\caption{Predicted single spin asymmetries for the process $p^\uparrow
p\to (K^+ + K^-)\,\, X$; kinematical conditions are the same as for the
pion case, at $p_T=1.5$ GeV/$c$. The curves correspond respectively
to the set of kaon FF's BKK1 \protect\cite{bkk1} (solid);
BKK2 \protect\cite{bkk2} (dashed); GR \protect\cite{gr2} (dot-dashed);
IMR \protect\cite{imr} (dotted). }
\label{kpm}
\end{figure}

In Fig.~\ref{kpm} we give and compare results obtained for $A_N(K^+ + K^-)$
with the different sets of fragmentation functions discussed above, 
i) to iv). This asymmetry is related to the asymmetries 
for the separate production of $K^+$ and $K^-$ by

\beq
  A_N(K^+ + K^-) = \frac{A_N(K^+) + \frac{\displaystyle 
  \sigma^{unp}(K^-)} {\displaystyle \sigma^{unp}(K^+)} \> A_N(K^-)}
  {1+\frac{\displaystyle \sigma^{unp}(K^-)}
  {\displaystyle \sigma^{unp}(K^+)}} \> \cdot
  \label{ak+k}
\eeq
According to Eq. (\ref{ak}) one would expect, at large $x_F$,  
$A_N(K^+ + K^-) \simeq A_N(K^+)$. All sets of FF yield similar results. 

Finally, we show in Fig. \ref{ks} the predictions obtained -- again using the 
different sets of fragmentation functions -- for $A_N(K^0_S)$. Contrary to 
the production of charged kaons, in this case the results strongly
depend on the set of fragmentation functions: sets which enhance the role, 
at large $z$, of valence quarks give results in agreement with the 
expectations of Eq. (\ref{ak}), whereas sets which allow a large
contribution from sea quarks give totally different results, even in sign. 
A measurement of $A_N(K^0_S)$ would then supply useful information. 

We have also investigated possible contributions from a non-zero
$I_{+-}$ distribution for the strange quark inside the polarized proton.
In principle, kaon asymmetries should be quite sensitive to this
distribution. However, as we have checked explicitely, at least in the 
kinematical region considered here, the role of strange quarks is strongly 
suppressed. This is because strange quarks come in any case
from the sea quarks in the proton, so that their (unpolarized and, to a greater
extent, polarized) distributions decrease at large $x$  much faster than
the distributions of $u$ and $d$ quarks.

The values of the SSA for kaons produced with polarized antiprotons,
$\bar p^\uparrow p \to K\,X$ can easily be deduced from those obtained with 
polarized protons by applying simple charge conjugation arguments, in analogy 
to Eq. (\ref{resppbar}); we have explicitely checked that this is the case. 

\section{Summary and conclusions}
\label{con}
We have consistently applied a QCD hard scattering scheme, with the inclusion
of some higher twist effects, to the description of single spin asymmetries
(SSA) in $\pup p \to \pi\,X$, refining a previous \cite{abm} determination of 
a new $\bfk_\perp$ and spin dependent distribution function, introduced by 
several authors \cite{siv1,siv2,abm,muld} as a possible source of single spin
asymmetries.

We have then used this distribution function to compute several single
spin asymmetries in other processes, namely $\bar p^\uparrow p \to \pi\,X$,
$\pup (\bar p^\uparrow) \, p \to \gamma\,X$ and $\pup (\bar p^\uparrow) \, p 
\to K\,X$. In the first two cases we have no free parameters and our results
are genuine predictions of the model: they turn out to agree with the
experimental data, although the data on SSA in $\gamma$ production are 
still rather qualitative with large errors. 

The SSA for the $\pup p \to K\,X$ process are predicted to be large, and, 
for neutral kaons, their actual value strongly depend on the set of 
fragmentation functions used; experimental information would allow to 
discriminate between different sets of kaon fragmentation functions. 
The main feature of the quark fragmentation which influences the value of 
the SSA is the relative importance of sea and valence quark contributions. 

Our scheme seems then to be a good phenomenological way of describing 
single spin asymmetries within a generalization of the QCD factorization
theorem; of course, other effects \cite{col1,tery}, might be present 
and play a smaller or larger role, depending on the process considered. 
For example, SSA in $\gamma$ production or Drell-Yan processes \cite{tery} 
should mostly originate from the mechanism used here or in Ref. \cite{ste1};
in other cases another or several other effects might be active at the 
same time. A possible strategy to isolate different origins of SSA in Deep
Inelastic Scattering has been discussed in Ref. \cite{alm}. More data and 
more phenomenological calculations are needed.

\acknowledgments

We are very grateful to M.E. Boglione, who contributed to the early stages
of this work. We wish also to thank M. Greco and S. Rolli; J. Binnewies, 
B.A. Kniehl and G. Kramer; D. Indumathi, H.S. Mani and A. Rastogi,
for kindly sending us a copy of their numerical
routines, and for clarifying discussions and comments on their
parametrizations of meson fragmentation functions.

\begin{figure}
\caption{
Predicted single spin asymmetries for the process $p^\uparrow p\to
K^0_S\,\, X$; kinematical conditions are the same as for the
pion case, at $p_T=1.5$ GeV/$c$. Notations for the theoretical curves are
the same as in Fig.~\protect\ref{kpm}. }
\label{ks}
\end{figure}

\clearpage

\begin{figure}[c]
\centerline{
\epsfig{figure=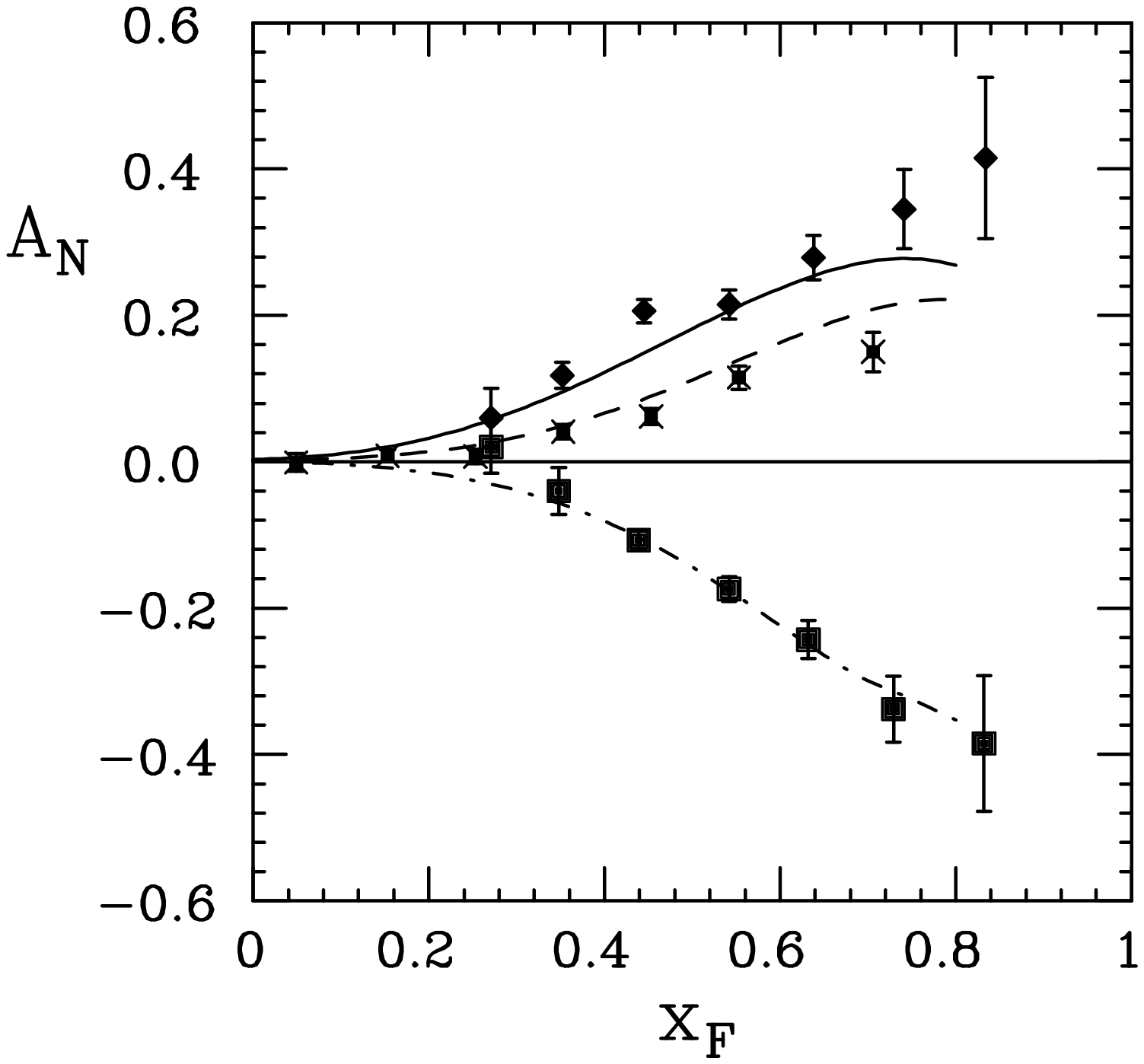,bbllx=50pt,bblly=200pt,bburx=530pt,%
bbury=650pt,width=15.0cm,height=15.0cm}}
 \begin{center}
 \begin{minipage}[c]{13cm}
 {\small {\bf Fig. 1:}
Fit of the data on $A_N$ for the process $p^\uparrow p\to\pi\,X$
\cite{exp1}, with the parameters given in Eq.~(\ref{par});
the upper, middle, and lower sets of data and curves refer
respectively to $\pi^+$, $\pi^0$, and $\pi^-$.}
 \end{minipage}
 \end{center}
\end{figure}

\clearpage

\begin{figure}[c]
\centerline{
\epsfig{figure=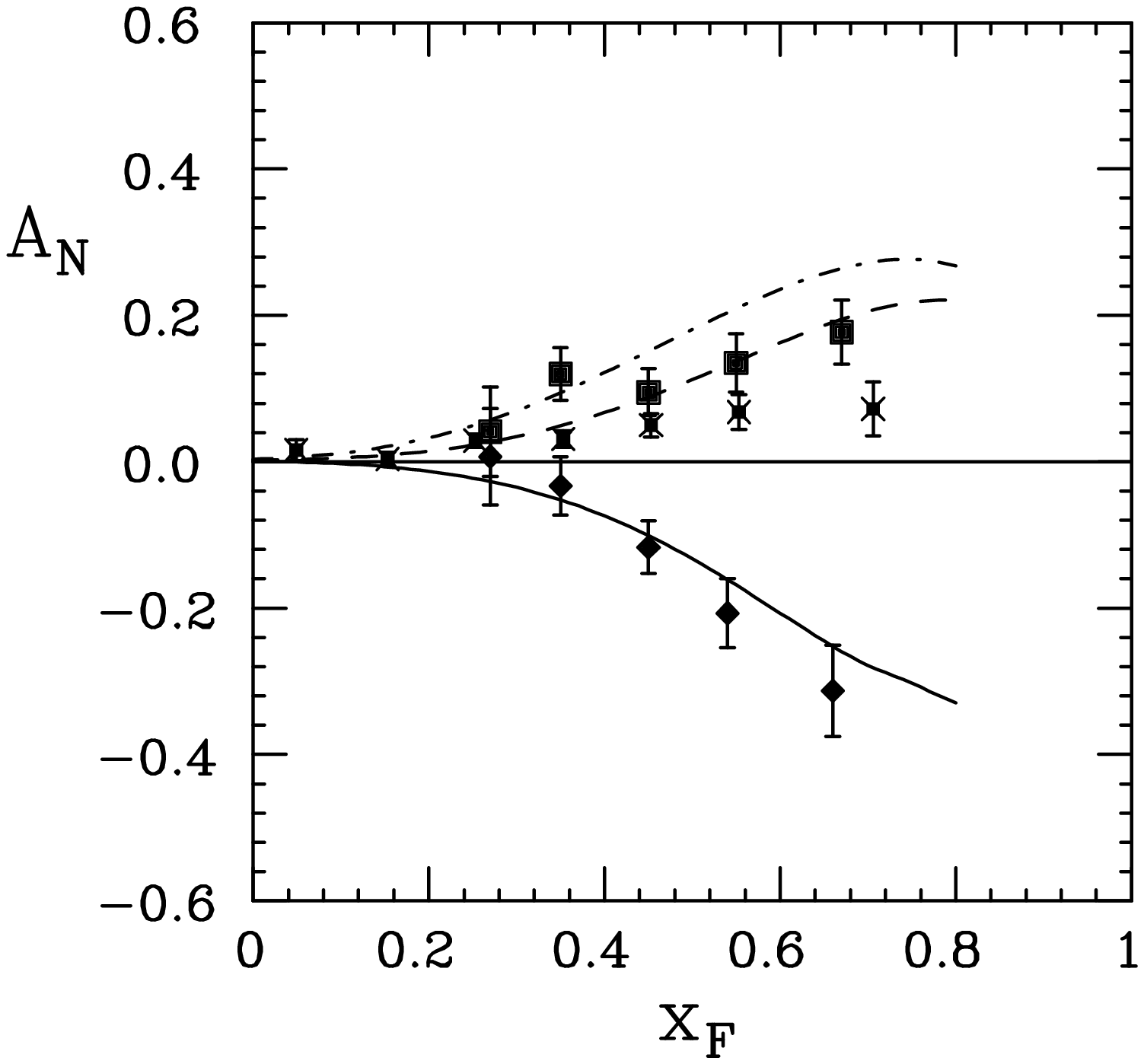,bbllx=50pt,bblly=200pt,bburx=530pt,%
bbury=650pt,width=15.0cm,height=15.0cm}}
 \begin{center}
 \begin{minipage}[c]{13cm}
 {\small {\bf Fig. 2:}
Single spin asymmetries $A_N$ for the process
$\bar{p}^\uparrow p\to\pi\,X$ \cite{exp2}; the lower, middle,
and upper sets of data and curves refer respectively
to $\pi^+$, $\pi^0$, and $\pi^-$.}
 \end{minipage}
 \end{center}
\end{figure}

\clearpage

\begin{figure}[c]
\centerline{
\epsfig{figure=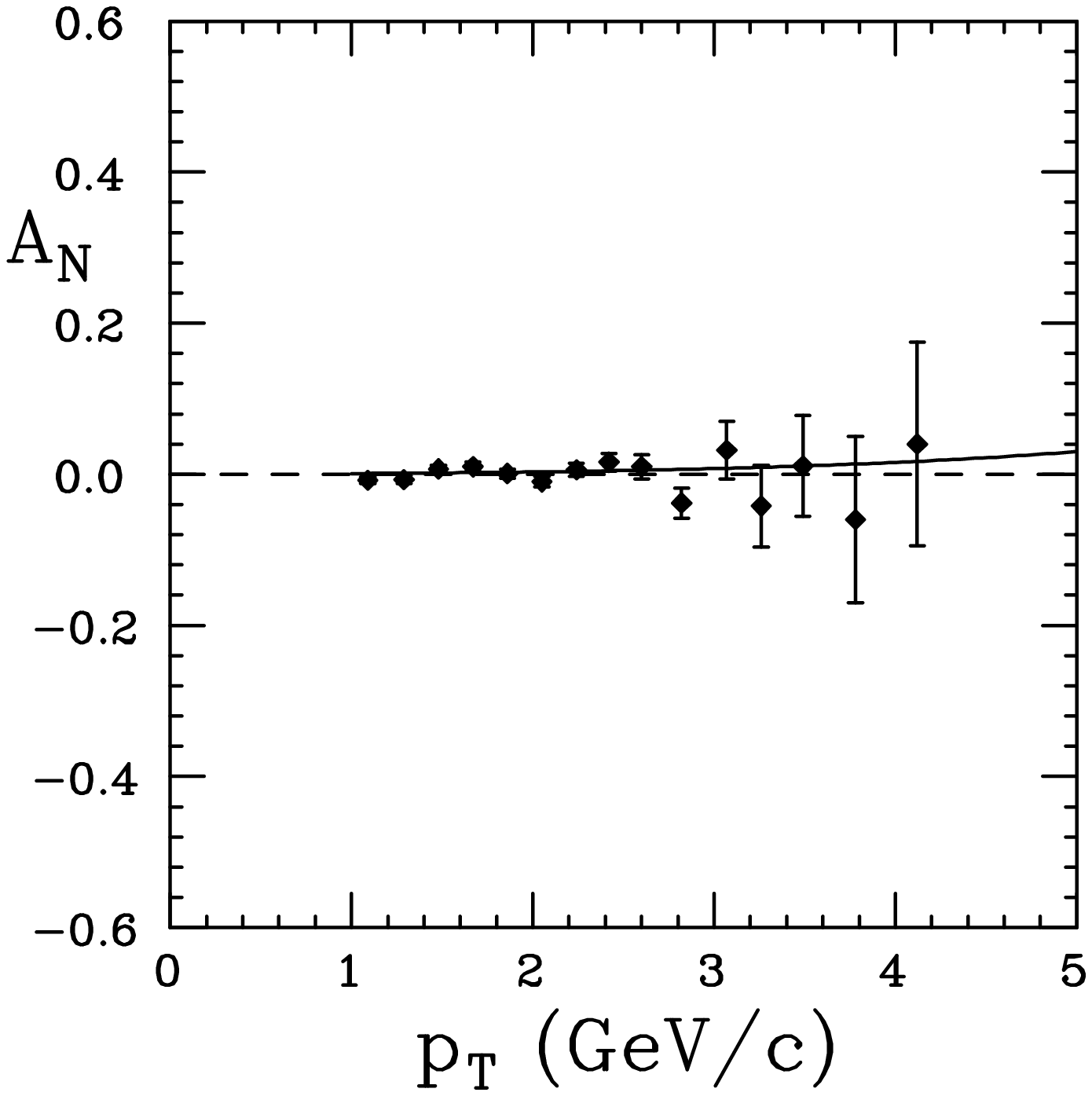,bbllx=50pt,bblly=200pt,bburx=530pt,%
bbury=650pt,width=15.0cm,height=15.0cm}}
 \begin{center}
 \begin{minipage}[c]{13cm}
 {\small {\bf Fig. 3:}
Single spin asymmetry for the process
$p^\uparrow p\to\pi^0\,X$ at fixed $x_F$, as a function of $p_T$;
experimental data, at $|x_F| \leq 0.15$, are from Ref.~\cite{exp3};
the solid curve shows our corresponding theoretical prediction at $x_F=0$.}
 \end{minipage}
 \end{center}
\end{figure}

\clearpage

\begin{figure}[c]
\centerline{
\epsfig{figure=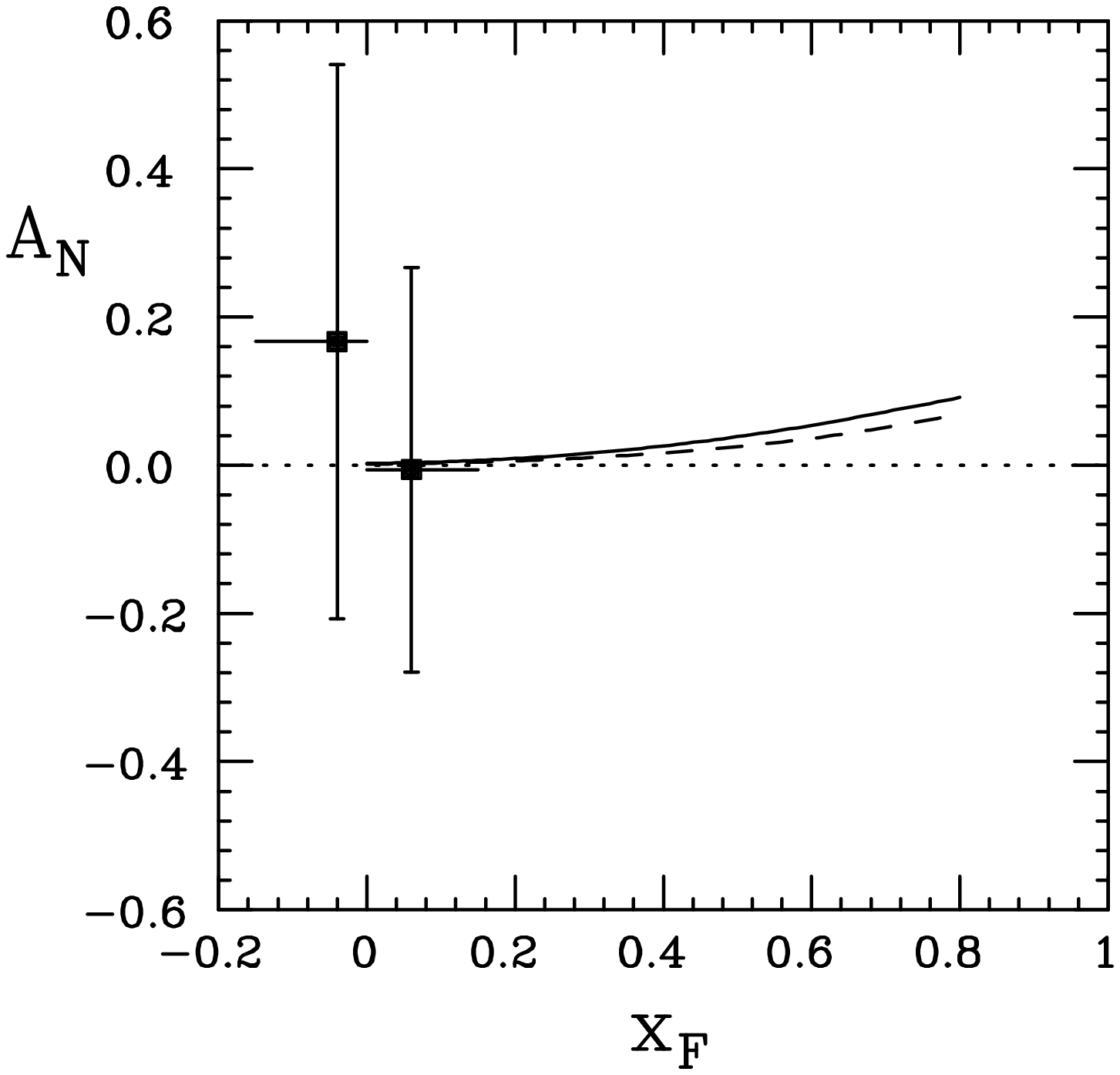,bbllx=50pt,bblly=200pt,bburx=530pt,%
bbury=650pt,width=15.0cm,height=15.0cm}}
 \begin{center}
 \begin{minipage}[c]{13cm}
 {\small {\bf Fig. 4:}
Single spin asymmetry for the process $p^\uparrow (\bar{p}^\uparrow) \,
p\to\gamma\,X$; experimental data, at $|x_F| \leq 0.15$ and 
$2.5 < p_T < 3.1$ GeV/$c$, are from Ref.\cite{exp4}. The curves show our 
corresponding theoretical predictions at $p_T=2.5$ GeV/$c$; the solid curve
refers to the $p^\uparrow p\to\gamma\,X$ process, the dashed curve to
the $\bar{p}^\uparrow p\to\gamma\,X$ case. }
 \end{minipage}
 \end{center}
\end{figure}

\clearpage

\begin{figure}[c]
\centerline{
\epsfig{figure=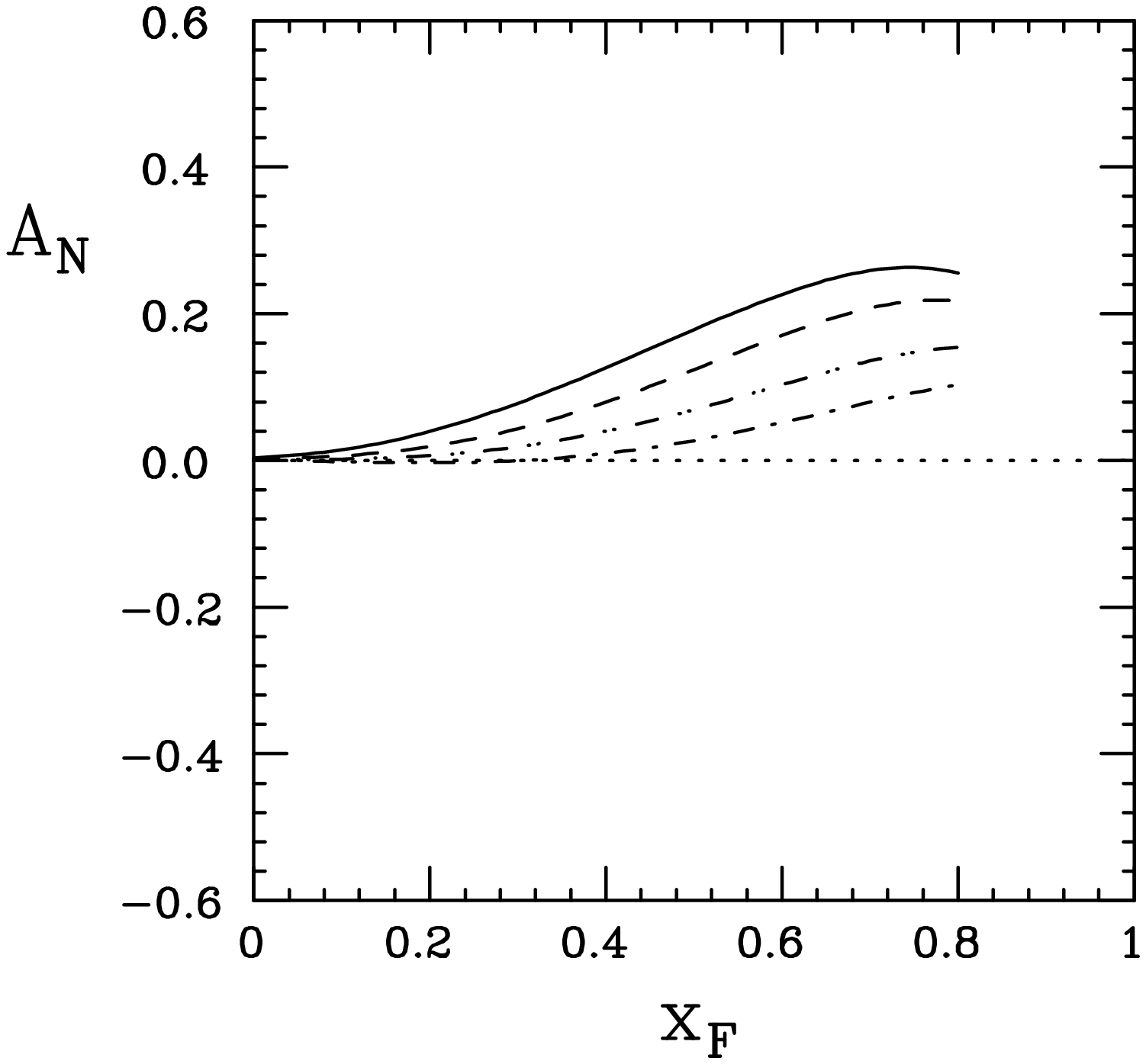,bbllx=50pt,bblly=200pt,bburx=530pt,%
bbury=650pt,width=15.0cm,height=15.0cm}}
 \begin{center}
 \begin{minipage}[c]{13cm}
 {\small {\bf Fig. 5:}
Predicted single spin asymmetries for the process $p^\uparrow p\to K\,X$,
with the set of kaon FF's BKK1 \protect\cite{bkk1}; kinematical conditions
are the same as for the pion case, at $p_T=1.5$ GeV/$c$. The solid,
dashed, dot-dashed, double dot-dashed curves refer respectively to
the $K^+$, $K^-$, $K^0$, $K^0_S$ cases. Results for $\bar{K}^0$ meson
are very similar to those for $K^-$ case. }
 \end{minipage}
 \end{center}
\end{figure}

\clearpage

\begin{figure}[c]
\centerline{
\epsfig{figure=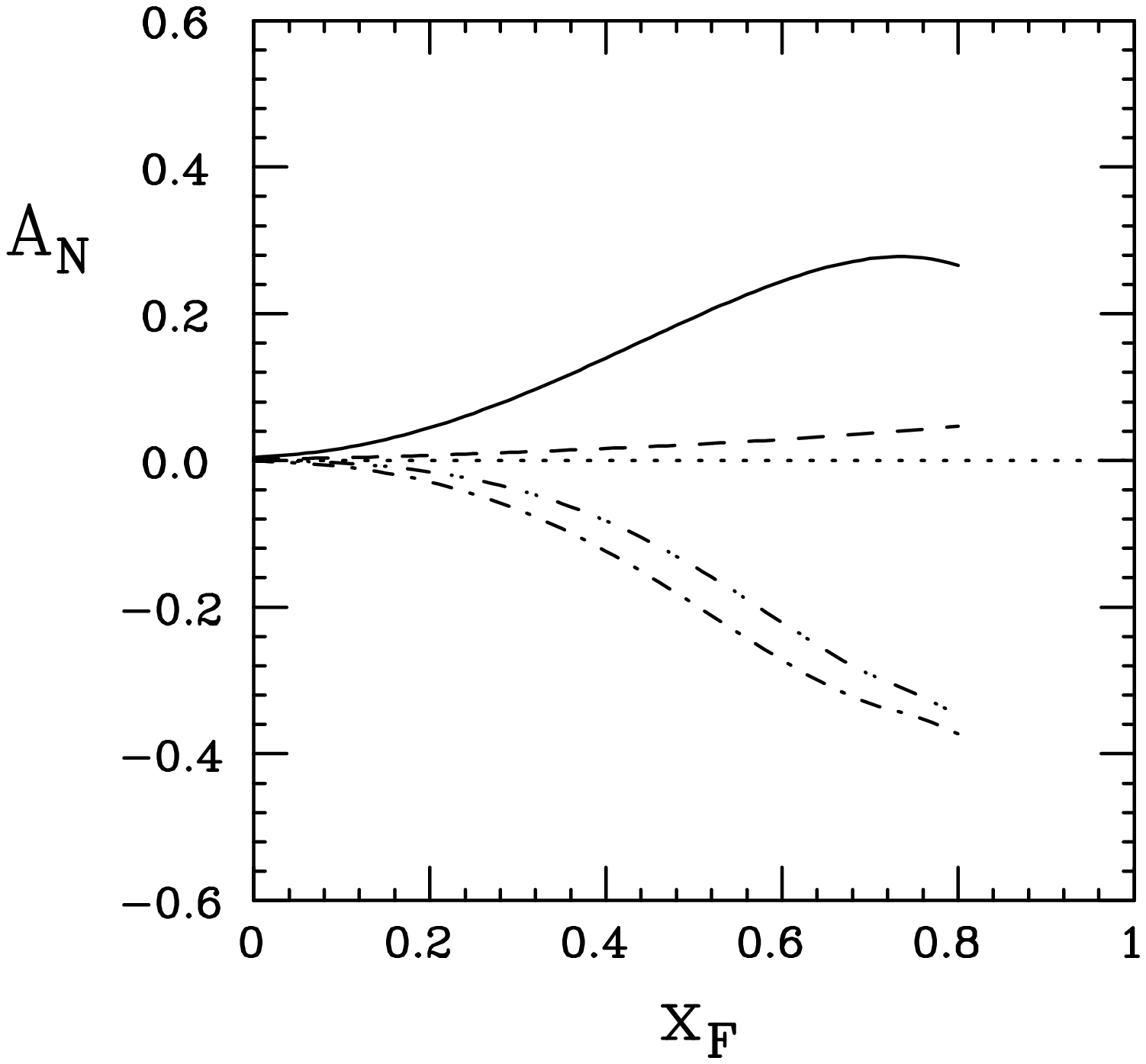,bbllx=50pt,bblly=200pt,bburx=530pt,%
bbury=650pt,width=15.0cm,height=15.0cm}}
 \begin{center}
 \begin{minipage}[c]{13cm}
 {\small {\bf Fig. 6:}
The same as for Fig.~\protect\ref{k1}, but using the set of kaon FF's
BKK1 modified so that $D_{K/sea}\simeq D_{K/val}[D_{\pi/sea}/D_{\pi/val}]$
(see text for more details). }
 \end{minipage}
 \end{center}
\end{figure}

\clearpage

\begin{figure}[c]
\centerline{
\epsfig{figure=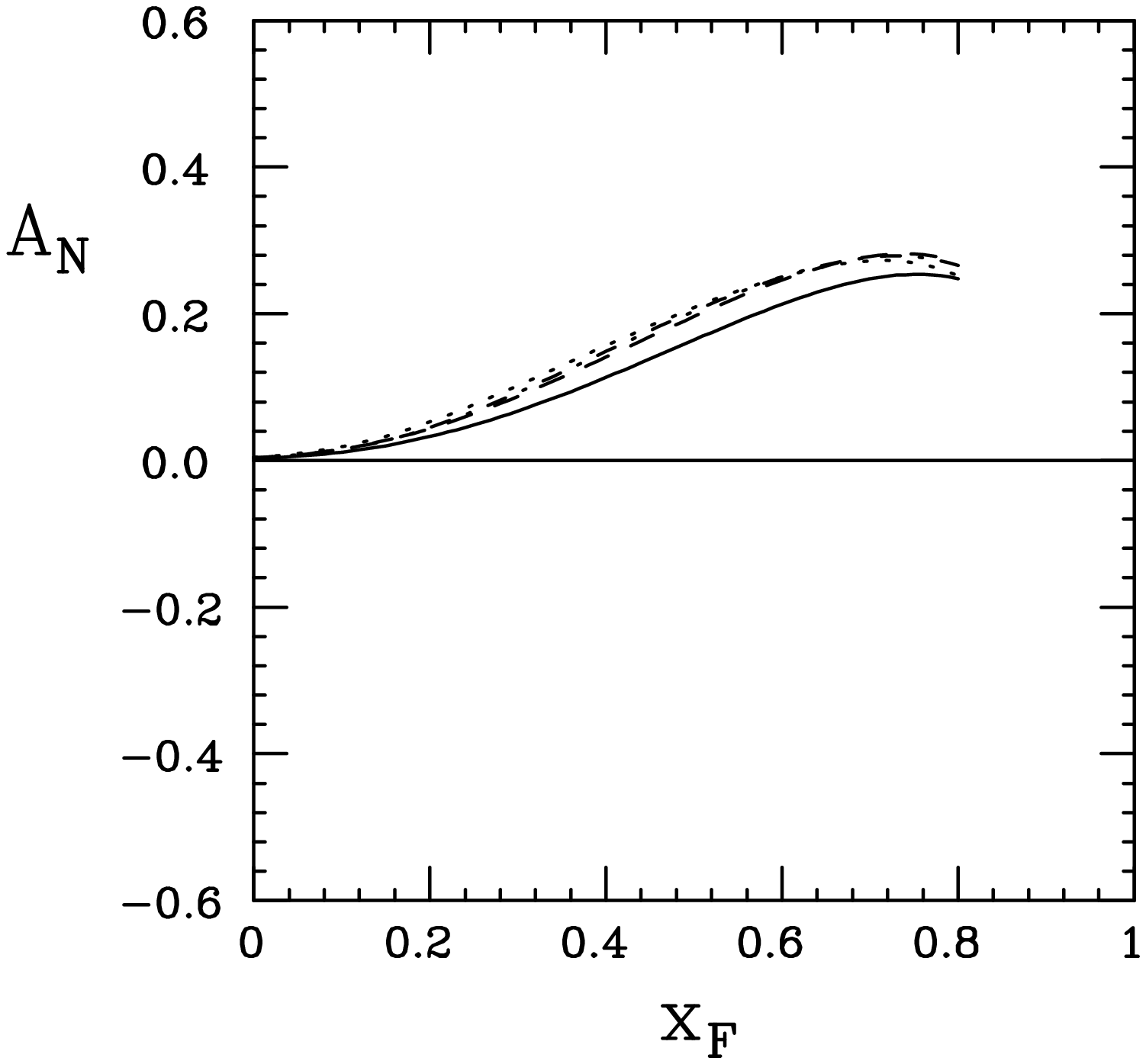,bbllx=50pt,bblly=200pt,bburx=530pt,%
bbury=650pt,width=15.0cm,height=15.0cm}}
 \begin{center}
 \begin{minipage}[c]{13cm}
 {\small {\bf Fig. 7:}
Predicted single spin asymmetries for the process $p^\uparrow p\to
(K^+ + K^-)\,\, X$; kinematical conditions are the same as for the
pion case, at $p_T=1.5$ GeV/$c$. The curves correspond respectively
to the set of kaon FF's BKK1 \protect\cite{bkk1} (solid);
BKK2 \protect\cite{bkk2} (dashed); GR \protect\cite{gr2} (dot-dashed);
IMR \protect\cite{imr} (dotted). }
 \end{minipage}
 \end{center}
\end{figure}

\clearpage

\begin{figure}[c]
\centerline{
\epsfig{figure=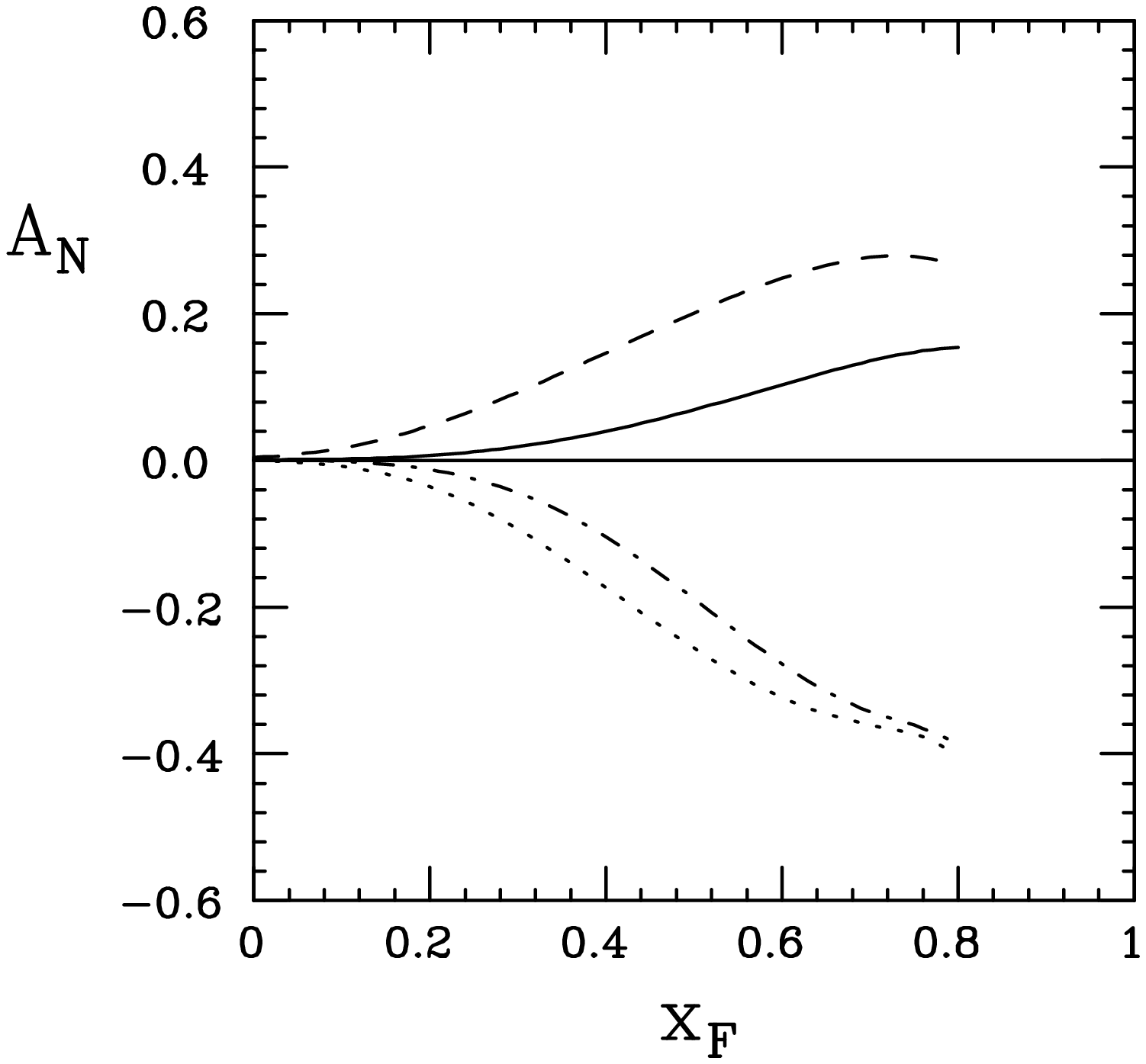,bbllx=50pt,bblly=200pt,bburx=530pt,%
bbury=650pt,width=15.0cm,height=15.0cm}}
 \begin{center}
 \begin{minipage}[c]{13cm}
 {\small {\bf Fig. 8:}
Predicted single spin asymmetries for the process $p^\uparrow p\to
K^0_S\,\, X$; kinematical conditions are the same as for the
pion case, at $p_T=1.5$ GeV/$c$. Notations for the theoretical curves are
the same as in Fig.~\protect\ref{kpm}. }
 \end{minipage}
 \end{center}
\end{figure}

\end{document}